\documentclass{article}

\usepackage{PRIMEarxiv}

\usepackage[utf8]{inputenc} 
\usepackage[T1]{fontenc}    
\usepackage{hyperref}       
\usepackage{url}            
\usepackage{booktabs}       
\usepackage{amsfonts}       
\usepackage{nicefrac}       
\usepackage{microtype}      
\usepackage{lipsum}
\usepackage{fancyhdr}       
\usepackage{graphicx}       
\usepackage{authblk}
\usepackage[backend=biber,style=ieee]{biblatex}

\title{AstroReview: An LLM-driven Multi-Agent Framework for Telescope Proposal Peer Review and Refinement}

\author[1,2]{Yutong Wang}
\author[1]{Yunxiang Xiao}
\author[1]{Yonglin Tian}
\author[1,2]{Junyong Li}
\author[1]{Jing Wang}
\author[1,2,*]{Yisheng Lv}

\affil[1]{The Key Laboratory of Multimodal Artificial Intelligence Systems, Institute of Automation, Chinese Academy of Sciences, Beijing 100190, China}
\affil[2]{The School of Artificial Intelligence, University of Chinese Academy of Sciences, Beijing 100049, China}

\begin{document}
\maketitle

\begin{center}
 \texttt{yutong.wang@ia.ac.cn}, \texttt{yunxiang.xiao@ia.ac.cn}, \texttt{yonglin.tian@ia.ac.cn},
 \texttt{lijy83@gmail.com},
 \texttt{wangjing2014@ia.ac.cn},
  \texttt{yisheng.lv@ia.ac.cn}
\end{center}
\vspace{2em}

\begin{abstract}
Competitive access to modern observatories has intensified as proposal volumes outpace available telescope time, making timely, consistent, and transparent peer review a critical bottleneck for the advancement of astronomy. Automating parts of this process is therefore both scientifically significant and operationally necessary to ensure fair allocation and reproducible decisions at scale. We present AstroReview, an open-source, agent-based framework that automates proposal review in three stages: (i) novelty and scientific merit, (ii) feasibility and expected yield, and (iii) meta-review and reliability verification. Task isolation and explicit reasoning traces curb hallucinations and improve transparency. Without any domain specific fine tuning, AstroReview used in our experiments only for the last stage, correctly identifies genuinely accepted proposals with an accuracy of 87\%. The AstroReview in Action module replicates the review and refinement loop; with its integrated Proposal Authoring Agent, the acceptance rate of revised drafts increases by 66\% after two iterations, showing that iterative feedback combined with automated meta-review and reliability verification delivers measurable quality gains. Together, these results point to a practical path toward scalable, auditable, and higher throughput proposal review for resource limited facilities.
\end{abstract}


\section{Introduction}

The effectiveness and quality of astronomical observations, especially those that target rapidly evolving transients, remain central to progress in modern astrophysics \cite{zhang2024grris, fang2025transientverse}. However, scarce telescope time combined with rising submission volumes places heavy demands on both applicants and review panels \cite{williams2025allocating}. These pressures create a persistent bottleneck in the observational pipeline.

For proposal authors, the entry barriers are high. Teams must navigate facility specific rules, evaluate target visibility and exposure time requirements, and substantiate novelty through comprehensive surveys of the literature and archival data. Preparation is further complicated by limited transparency: most observatories release only the abstracts of successful proposals, whereas rejected submissions and referee reports remain confidential. Without access to negative exemplars and standardized optimization guidance, applicants frequently rely on iterative resubmissions with uncertain payoffs. From the reviewers’ standpoint, the workload is equally formidable. Time Allocation Committees (TAC) often assess hundreds to thousands of proposals per cycle under tight deadlines. Reviewers must judge scientific merit, technical feasibility, and related criteria, and they must reconcile scores during panel deliberations within a short window. Escalating submission rates have amplified concerns about reviewer fatigue, inconsistent grading, and the sustainability of current practices. 

Recent studies have recognized the increasing complexity and time-consuming nature of the proposal review process, particularly in large-scale scientific facilities \cite{kerzendorf2020distributed, carpenter2025enhancing}. To address these challenges, several efforts have explored the application of Artificial Intelligence (AI) techniques to support reviewer assignment, primarily focusing on optimizing the match between proposals and reviewers based on topical similarity or inferred expertise. However, the applicability of these approaches remains limited because they regard AI primarily as a pre‑processing tool for allocating review tasks, rather than as a means of providing direct assistance during the evaluation itself, by assessing proposal quality or offering constructive feedback.

As LLMs \cite{hurst2024gpt, grattafiori2024llama} continue to demonstrate exceptional reasoning abilities across a variety of tasks, researchers are increasingly focusing on fine-tuning these models with domain-specific data for use in academic peer review. Recent studies have shown that LLMs can achieve review accuracy comparable to that of human experts \cite{lu2024ai}.
Inspired by this, we present AstroReview, a cooperative agent‑based review architecture that acts as a controlled assistive tool for observing proposal review. Built upon open‑source large language models serving as autonomous agents, the system emulates expert panels and conducts a three‑stage assessment: (i) novelty and scientific merit, (ii) feasibility and expected yield, and (iii) meta-review and reliability verification, thus delivering comprehensive, fine‑grained comments and recommendations. While the proposed methodology is tailored to the evaluation of astronomical observing proposals, its overarching review pipeline can be readily adapted to proposal assessments in other scientific and engineering domains.

Our key contributions are summarized as follows:
\begin{itemize}
    \item Inspired by the step‑wise structure of real‑world review workflows, AstroReview reproduces this natural reasoning sequence and confines each agent to a single, clearly defined task at every stage. By partitioning the evaluation pipeline into isolated phases, the framework localizes potential reasoning errors, reduces large‑model hallucinations and preference drift, and delivers clearer, more trustworthy review comments and recommendations.
    \item Even without any domain-specific few-shot prompts or fine-tuning, two key experiments validate the effectiveness of the meta-review and reliability verification stage. The Review Agent correctly identifies genuinely accepted proposals with an accuracy of 87\%. Once drafts pass through the two-round review-refinement loop, the Proposal Authoring Agent lifts the overall acceptance rate by 66\%, demonstrating that iterative feedback yields substantial, measurable gains across every quantitative metric.
    
\end{itemize}

\section{Related Work}

\subsection{LLMs in Paper Review}

As LLMs increasingly demonstrate expert-level capabilities in evaluation \cite{zheng2023judging}, growing attention has been directed toward their application in the domain of peer review \cite{yu2024automated, idahl2024openreviewer}. A range of recent studies has explored how to leverage LLMs to improve the peer review process, primarily through few-shot prompting, self-reflection, and supervised fine-tuning techniques. With the emergence of agent-based frameworks, researchers have begun to shift focus to incorporating LLMs into structured peer-review workflows involving multiple reviewers, meta-review stages, iterative review-refinement cycles, and decomposed sub-tasks \cite{lu2024ai, d2024marg, yamada2025ai, weng2025cycleresearcher, zhu2025deepreview}, leading to more rigorous and insightful peer review outcomes. However, a European Southern Observatory (ESO) pilot study directly applied ChatGPT‑3.5/4 to score and rewrite five real observatory proposals, concluding that generic LLMs provide little benefit for either automatic scoring or full‑text polishing \cite{jerabkova2024scientific}.

\subsection{LLMs in Astronomy}

The formidable knowledge-acquisition and reasoning capabilities of Transformer\-based large language models have catalyzed a rapid surge of astronomy-specific applications \cite{li2025astronomical}. AstroBERT \cite{grezes2021building} first adapted BERT \cite{devlin2019bert} to about 400k documents from the Astrophysics Data System (ADS), enabling deep contextual understanding and entity recognition. AstroLLaMA \cite{nguyen2023astrollama, perkowski2024astrollama} leveraged LLaMA-2 backbones \cite{touvron2023llama} and 300k astronomy abstracts from arXiv to deliver generative Question Answering (QA). The community-driven AstroMLab introduces AstroSage-Llama-3.1-70B \cite{de2025astromlab}, an astronomy-specialized LLM trained with continued pre-training, supervised fine-tuning, and model merging, which attains 86.2\% accuracy on the AstroMLab-1 benchmark \cite{ting2025astromlab}, surpassing both proprietary \cite{pan2024astromlab, deastromlab} and general models, as well as human expert performance on this task. However, excessive infusion of astronomy‑specific knowledge during fine-tuning can trigger catastrophic forgetting, severely eroding the model’s general knowledge base and sharply degrading its reasoning and instruction-following abilities, thereby making it unsuitable to serve as an interactive agent in proposal review.

\section{AstroReview: A Multi-Agent Framework for Proposal Review}

\begin{figure}
	\centering
	\includegraphics[width=\linewidth]{Figs/AstroReview.pdf}
	\caption{Overview of AstroReview. The framework emulates human expert evaluation across three key dimensions: novelty and scientific merit, feasibility and expected yield, and meta-review and reliability verification.}
	\label{fig:AstroReview}
\end{figure}

The LLM-driven multi-agent astronomical assistant framework is illustrated in Figure \ref{fig:AstroReview}, which proceeds through three consecutive stages.

\textbf{Stage 1: Assessment of Novelty and Scientific Merit.} A semantic parser first extracts key objects, phenomena, and research goals from the proposal draft. Guided by these cues, the question‑generation module crafts precise evaluative queries to test whether the proposed observations of the selected objects truly extend current understanding and deliver substantive scientific value. For instance, by asking if analogous observations have already been carried out with the same or a more sensitive instrument, whether existing archives already contain sufficient data to address the stated objectives, or whether the proposed cadence captures information unavailable from prior surveys. These queries inform a retrieve-and-rank process conducted across scientific literature, web resources, and curated databases to gather relevant evidence. Finally, a novelty assessment module integrates the retrieved evidence to synthesize an expert-style judgment on the proposal’s originality and potential scientific contribution.

\textbf{Stage 2: Assessment of Feasibility and Expected Yield.} A parameter parser extracts instrument settings, orbital constraints, and other scheduling details directly from the proposal. Through calls to the mission’s official simulation tools, these parameters drive two complementary analyses: Visibility prediction determines accessible windows and per‑orbit efficiency for each target, while observation simulation generates synthetic exposures using the proposed integration times, yielding signal‑to‑noise diagnostics and preview frames. Together, the results provide a quantitative assessment of the plan’s practical feasibility and expected scientific yield.

\textbf{Stage 3: Meta-Review and Reliability Verification.} Stage 3 ingests the full proposal together with the structured novelty and feasibility assessments produced in Stages 1 and 2. It evaluates the proposal against each criterion and produces detailed, criterion‑referenced critiques. Multiple independent evaluations are then consolidated through a meta‑review that issues a final decision accompanied by succinct summary comments. This consolidation mitigates stochastic variation and individual bias, yielding a stable, internally consistent verdict. Finally, reliability verification is carried out through an external audit routine that enforces template compliance, cross‑checks every claim against its supporting evidence, and immediately revises any hallucinations or logical inconsistencies it uncovers.

By separating novelty assessment, technical feasibility checks, and consensus-driven scoring into sequential yet interlinked stages, the framework isolates reasoning faults, suppresses hallucinations, and produces traceable, fine-grained feedback well suited to proposal evaluation.

\section{AstroReview in Action: Iterative Review-Refinement Cycles}

\begin{figure}[tb]
	\centering
	\includegraphics[width=\linewidth]{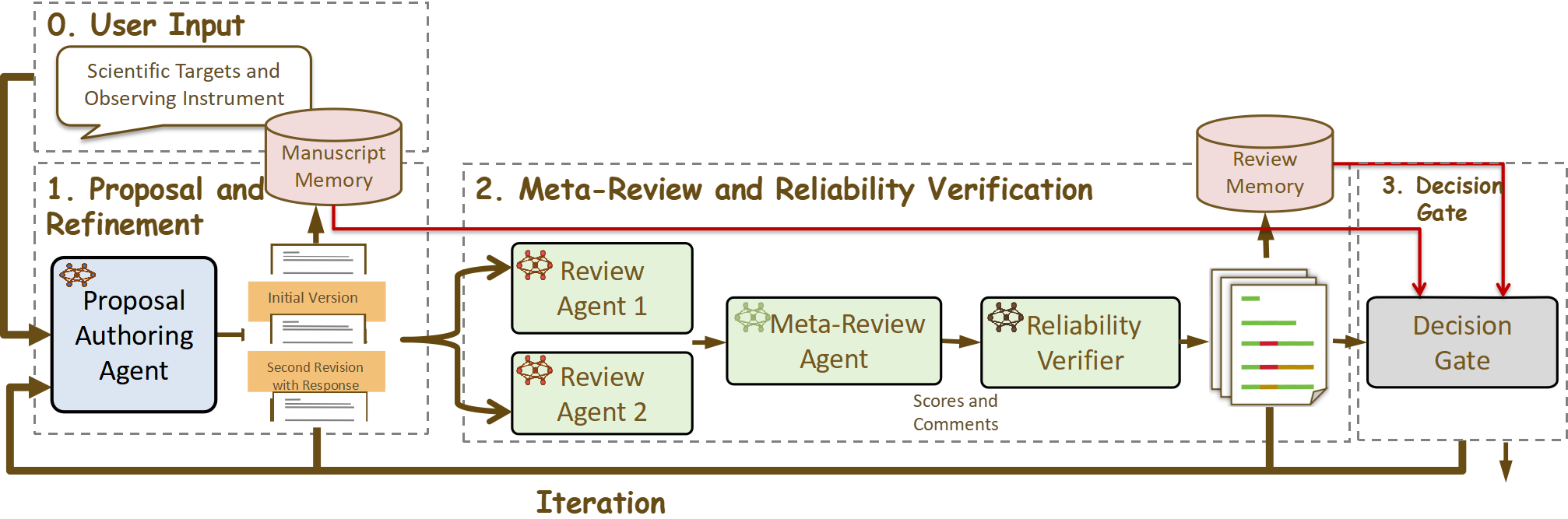}
	\caption{Workflow of the iterative proposal review and refinement. The process starts with user inputs that specify scientific targets or phenomena and the observing instrument (Step 0). In Step 1 (Proposal and Refinement), the Proposal Authoring Agent drafts or revises the proposal using prior feedback stored in the Manuscript Memory. In Step 2 (Review), multiple Review Agents generate scores and comments, which are aggregated by the Meta-Review Agent and then audited by the Reliability Verifier for template compliance, logical coherence, and evidence alignment; the resulting records are stored in the Review Memory. In Step 3 (Decision Gate), the system decides whether to stop or continue another iteration based on the proposal content and reviewer feedback. The loop repeats until the stopping condition is satisfied.}

	\label{Iteration}
\end{figure}

To assess AstroReview’s effectiveness in proposal evaluation, we establish an iterative optimization loop that feeds AstroReview’s review comments and actionable suggestions back into the proposal-drafting process, thereby simulating the typical review-refinement cycle. As illustrated in Figure~\ref{Iteration}, the workflow is driven by two core components: a Proposal Authoring Agent, responsible for scientific writing and revision, and a Review Agent, which evaluates each draft. Together, they form a closed loop in which successive rounds of critique progressively steer the proposal toward a higher-quality, more competitive version. 

Publicly available corpora of complete observing proposals are exceedingly scarce. Consequently, our experiments rely exclusively on the Proposal Abstracts Catalog, which contains only the abstracts of successful Hubble Space Telescope (HST) submissions. An abstract, however, is merely a concise excerpt of the full proposal, and the catalog’s entries vary widely in writing style, formatting, length, etc. This heterogeneity yields a fragmented linguistic signal that prevents the semantic‑ and parameter‑parsing modules in Stages 1 and 2 from reliably extracting the structured information needed for novelty and feasibility assessment. Accordingly, the current implementation of the Review Agent bypasses these two stages and operates solely at Stage 3.

\subsection{Proposal Authoring Agent}

The Proposal Authoring Agent functions as a scientific writer. Guided by a predefined structural template, it iteratively revises the draft and composes responses to reviewer feedback, weaving suggested changes into successive proposal versions. Because these AI‑generated drafts are never intended for direct submission, we intentionally refrain from dissecting or optimizing the agent’s internal planning and task-execution routines. Nonetheless, the Proposal Authoring Agent remains essential for showcasing the Review Agent’s ability to deliver constructive, end-to-end support throughout the proposal-development cycle.

\subsection{Review Agent}

The Review Agent is seeded with a system prompt directing it to emulate several independent reviewers and to record their assessments in a predefined structured template. Each simulated reviewer scores the proposal on six dimensions: (i) Scientific Impact, (ii) Scientific Breadth, (iii) Methodological Innovation and Uniqueness, (iv) Facility Necessity, (v) Technical Feasibility, and (vi) Confidence in Deliverable Outcomes. For every critical comment, the reviewer must provide (a) the exact passage that triggers concern, (b) a concise diagnostic explanation, and (c) a concrete revision recommendation. A meta-review then synthesizes the individual reports, issuing a definitive decision accompanied by a succinct panel summary. Subsequently, an external audit module audits the entire output for hallucinations, logical inconsistencies, and formatting deviations, automatically rectifying any detected deficiencies.

The meta‑review stage reconciles the individual reviewer reports, reduces individual bias and model‑induced stochastic bias, promoting convergence on a stable verdict. The additional reliability verification sweep further suppresses LLM hallucinations and reasoning errors, thereby enhancing the overall reliability of the evaluation pipeline. Additionally, experiments reveal that when a reviewer fails to cite the precise locus of a defect, the Proposal Authoring Agent is prone to stochastic edits and hallucinations, often modifying unrelated sections and occasionally altering text that was originally correct. Accordingly, the reviewer instructions were refined to mandate explicit passage citation, which markedly improves the precision and relevance of subsequent revisions.

\subsection{Decision Gate}

Within the review-refinement loop, a dedicated Decision Gate module determines whether additional iterations are warranted. At the conclusion of each cycle, the module evaluates two signals: (i) Textual similarity between the current and preceding drafts. If the cosine similarity exceeds 0.90, the revision is considered insubstantial. (ii) Score convergence, measured as the absolute difference between consecutive scores. An improvement of $<$ 1 point is interpreted as stagnation. When both conditions are satisfied, or when the predefined maximum of three iterations has been reached, the module asserts a stop flag. By continuously monitoring progress and identifying diminishing returns, it ensures early termination of the loop, thereby preventing superfluous computational expenditure while preserving review quality.

Both termination criteria are indispensable, because relying on either one alone may allow stochastic fluctuations in the LLM to halt the optimization loop prematurely. The textual-similarity threshold is included to counteract what we have empirically observed as ``model inertia'': during multi-round revisions the generator sometimes responds to reviewer feedback with only superficial, near-duplicate edits, and conversely the reviewer may re-issue essentially the same comments. By measuring overlap we force substantive changes before the process can finish. The score-convergence test addresses the complementary failure mode in which the reviewer’s numeric assessment remains unchanged from one round to the next merely because of sampling variance rather than genuine convergence. Used together, these safeguards ensure that the loop continues until the proposal is meaningfully improved and not merely cycling through random or trivial adjustments.

\subsection{Memory Module}

To enable the Decision Gate’s similarity computation, we maintain two separate memory buffers, one for proposal drafts and one for reviewer reports, throughout the dialogue between the Proposal Authoring Agent and the Review Agent. During the iterative experiments, we deliberately disabled the agent’s native memory module and instead manually injected the two most recent memory buffers into the Proposal Authoring Agent and Review Agent. We adopt this strategy because, in multi‑round refinement, relying on the agent’s own memory often pushes the combined length of memories, prompts, and expected outputs beyond the model’s maximum context window, resulting in truncated input semantics or outputs that are too brief to meet our requirements.

\subsection{Reliability Verification}

Empirical experiments reveal the indispensability of a reliability verification module. Because stochastic sampling and the finite reasoning capacity of LLMs often lead to schema non-compliance or internal inconsistencies in the generated reviews, the reliability verifier conducts a post-hoc audit to identify and amend such defects. Although this mechanism cannot eradicate all errors, it markedly enhances the reliability of the final evaluation output, thereby increasing confidence in the review process.

\section{Experiments}

\subsection{Dataset Construction}

We use accepted HST proposal abstracts as positive samples. Since most observatories, including HST, do not publicly release rejected proposals, we generate negative samples by applying controlled perturbations to the accepted proposals according to predefined rules. Currently, there are 13411 accepted HST proposals available. To address the class imbalance issue, we construct a balanced dataset with an equal number of positive and negative samples (1:1 ratio), resulting in a total of 26,822 proposals. This ensures that the evaluation of classification performance is not biased toward the majority class and allows for a fair comparison across metrics.

We generate negative samples by systematically degrading each accepted abstract until it resembles the kind of unfocused, unconvincing text that is typically rejected by review panels. The transformation begins by randomly deleting a proportion of sentences, which immediately breaks the narrative continuity and leaves key arguments under-developed. The surviving sentences are then shuffled, further scrambling the logical flow and making it difficult for a reader to trace the thread of reasoning. Next, we blur technical precision: concrete numbers, dates and instrument names are replaced with generic stand-ins such as “some,” “20XX” or “a space telescope,” while hedging phrases like “possibly” and “in theory” are injected to undermine the author’s confidence. Finally, an optional truncation step lops off the tail of the abstract, creating abrupt endings and information gaps. By tuning the deletion rate, vagueness probability and truncation rate, we control how severely the text quality is diminished, ensuring the negatives remain grammatically sound and on topic but clearly fall short of the standards that earned the originals their acceptance labels. Specifically, the deletion rate, vagueness probability, and truncation rate are set to 0.35, 0.25, and 0.15, respectively.

To prevent the evaluator from being influenced by metadata such as Prop. Type, Category, ID, Cycle, Title, and PI, which may lead to label leakage, we perform a de-identification process. Specifically, we remove all such metadata from the HST proposal abstracts before feeding them into the evaluator, ensuring that the evaluation is based solely on the scientific content of the abstract.

\subsection{Evaluation Metrics}

Based on the constructed dataset, we conduct two types of experiments to evaluate the effectiveness of the Review Agent.

\textbf{Decision Evaluation} The first evaluation focuses on the accuracy of the review decisions. To verify whether the Review Agent can distinguish between accepted (positive) and rejected (negative) proposals, we structure its outputs as a binary classification task. The classification performance is assessed using standard metrics including F1-score, Accuracy, Precision, and Recall, measuring the correctness of its decisions.

\textbf{Proposal Improvement Evaluation} This second evaluation stage measures how effectively the Review Agent’s feedback drives proposal refinement. Effectiveness is quantified by two metrics: (i) the incremental increase in scores and (ii) the corresponding rise in proposal acceptance rate.

\subsection{Experimental Settings}

We utilize the locally deployed Qwen-2.5-72B \cite{qwen2024qwen25} via Ollama for both Proposal Authoring Agent and Review Agent, since it has the best performance among open-weights models. It demonstrates significantly improved reasoning ability and instruction-following performance aligned with human preferences. It also shows enhanced capabilities in understanding and generating structured data formats. During inference, we fix the sampling temperature at 0.4 and cap both the input and output sequence lengths at 100k tokens. Inference is performed on an NVIDIA H100 GPU.

Each agent, Proposal Authoring Agent, Review Agent, and Reliability Verifier, is driven by a dedicated system prompt. The Review Agent prompt is designed to emulate a human proposal reviewer: it assigns scores to the proposal abstract, formulates clarifying questions and constructive feedback, and ultimately issues an accept-or-reject recommendation. Although the evaluation rubric embedded in the prompt differs in detail from the official criteria adopted by some observatories, it intentionally adheres to a domain-agnostic framework. Specifically, we assess each proposal based on six key dimensions: (i) Scientific Impact, (ii) Scientific Breadth, (iii) Methodological Innovation and Uniqueness, (iv) Facility Necessity, (v) Technical Feasibility, and (vi) Confidence in Deliverable Outcomes. This design choice aims to ensure broad applicability across evaluation processes at different observatories.

\subsection{Decision‑Making Accuracy of the Review Agent}

\begin{table}
    \centering
    \caption{Ablation study on the Review Agent powered by Qwen‑2.5‑72B.}

    \begin{tabular}{p{5cm} cc}
    \toprule
    Category  & Positive Accuracy & Negative Accuracy\\
    \midrule
    Single Reviewer & 79.78\% & 58.46\% \\
    Multi-Reviewer + Meta-Review & 80.51\% & 48.16\%\\
    Multi-Reviewer + Meta-Review + Reliability & \textbf{87.08\%} & 46.69\% \\
    \bottomrule
    \end{tabular}
    \label{tab:statistics}
\end{table}

Ablation results (Table~\ref{tab:statistics}) quantify the Review Agent’s ability to classify proposal abstracts. We employ two Review Agents in the multidimensional reviewing stage. Preliminary experiments showed that adding more agents yields highly redundant comments with limited marginal utility, which increases computational cost and, more importantly, reduces opinion diversity for meta-review synthesis. Meta-review benefits from complementary perspectives rather than multiple restatements of the same critiques. It is crucial to report accuracy separately for positive and negative samples. During our experiments we noticed that the “negative” drafts we synthesize still retain features that appeal to reviewers, so in the ablation study our strongest configuration, multi-reviewer combined with meta-review and reliability verification, raises the accuracy on true positives yet drives accuracy on negatives down to only 46\%, actually worse than random guessing (50\%). The so-called rejected cases are intentionally degraded versions of proposals that were once accepted; even after the prose is diluted, the drafts still present attractive targets and significant scientific value, making them hard for the agent to dismiss. Therefore, evaluating the Review Agent on proposals that were truly rejected by a TAC would provide a more accurate and meaningful measure of its real-world performance.

As expected, the single‑reviewer baseline attains the lowest positive accuracy. Adding multiple reviewers and feeding their reports to a meta-reviewer lifts that accuracy only slightly, because the meta-reviewer’s synthesis frequently diverges from individual assessments. Accuracy peaks only after a reliability verification stage is added, confirming that the full stack, multiple reviewers, a meta-review, and reliability verification, works synergistically to flag bias, curb hallucinations, and surface logical inconsistencies.

\subsection{Proposal Improvement Guided by the Review Agent}

\begin{figure}
	\centering
	\includegraphics[width=.8\linewidth]{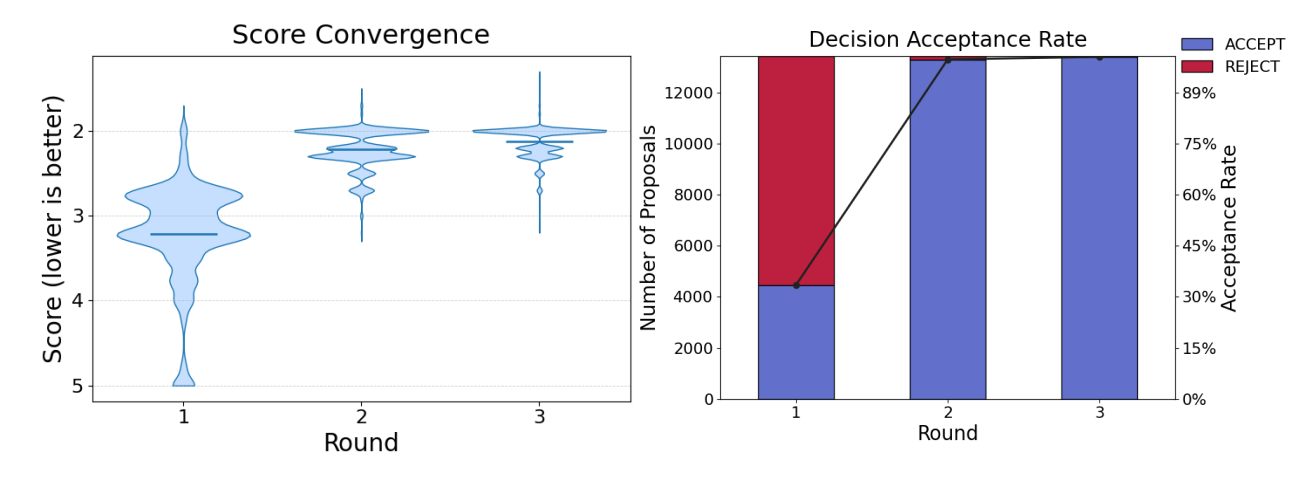}
	\caption{Evolution of proposal evaluation metrics across three refinement rounds. Violin plots depicting the full distribution of scores in each round (left); Bar chart summarizing the acceptance and rejection rate achieved in each round (right).}
	\label{fig:results}
\end{figure}

We evaluate reviewer scores and acceptance rate over three refinement rounds, using a single reviewer per round to limit computational overhead. Results are summarized in Fig.~\ref{fig:results}. During the first revision the mean score fell by 0.99 points, while the standard deviation contracted from 0.59 to 0.23. This steep drop, accompanied by tighter dispersion, shows that the Review Agent’s initial comments corrected fundamental weaknesses, scientific focus, feasibility rationale, and alignment with the review rubric, thereby lifting proposals to a more uniform quality level. The second revision produced only a further 0.09 point decline, suggesting diminishing returns once major structural flaws had been removed; at that stage the feedback mostly refined wording and technical detail rather than inducing additional conceptual change. Acceptance dynamics echo this pattern: probability of acceptance jumped from roughly 33\% to 99\% after the first revision, then inched up by just 0.75\% in the final pass.

The near‑perfect figure in Round 3 reveals a ceiling effect, implying that the corpus had essentially reached the acceptance threshold and that the final comments provided polish rather than rescuing borderline cases. Taken together, these trends indicate that high-impact, early feedback loops are the engine of substantive improvement, whereas later cycles serve chiefly for bespoke fine‑tuning. Overall, the data demonstrate that the Review Agent’s guidance was decisive in producing substantive improvements.

\subsection{Qualitative Assessment of Review Agent Competence}

\begin{figure}
	\centering
	\includegraphics[width=\linewidth]{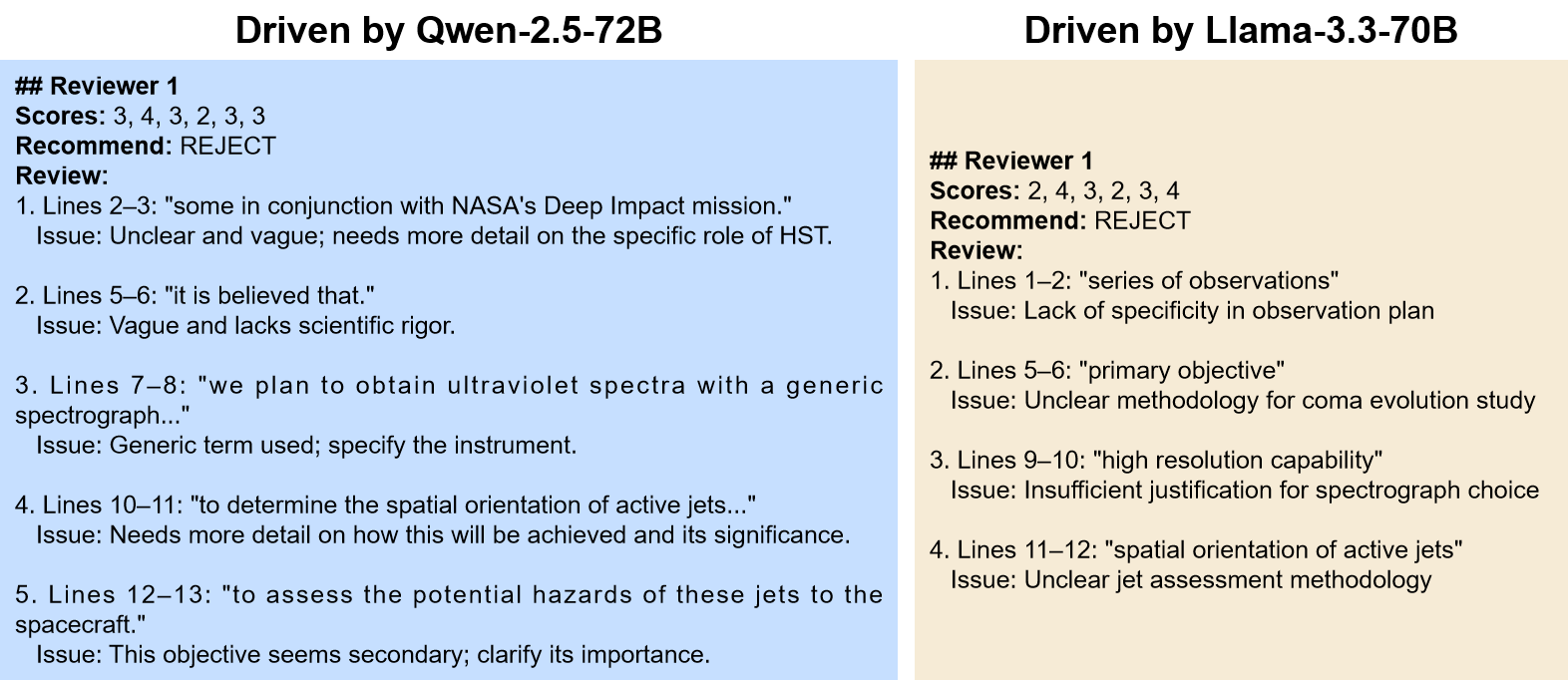}
	\caption{Representative Review Agent outputs generated using Qwen‑2.5‑72B and Llama‑3.3-70B back‑ends.}
	\label{Response}
\end{figure}

Figure~\ref{Response} displays representative excerpts generated by the Review Agent when powered by Qwen‑2.5‑72B and Llama‑3.3‑70B\cite{grattafiori2024llama}. The discussion below addresses (i) model selection and (ii) qualitative review performance. Extensive testing revealed that sub‑70‑billion‑parameter models (e.g., 32B and 8B) are ill‑suited to proposal review: their outputs are frequently malformed, score-decision alignment is poor, and they fail to localize defective passages in rejected abstracts. These issues stem from limited long‑range reasoning and weaker instruction following, rendering this parameter tier inadequate.

We therefore benchmarked three state‑of‑the‑art 70‑billion‑parameter LLMs, Qwen‑2.5‑72B, Llama‑3.3‑70B, and DeepSeek‑R1‑70B\cite{deepseekai2025deepseekr1incentivizingreasoningcapability}. Qwen‑2.5‑72B adhered flawlessly to the prescribed template and showed the strongest alignment between its numerical scores and final recommendations. Its grading rubric was applied with notable rigor and caution, and every issue it highlighted matched the deliberately perturbed passages exactly, producing granular, substantive commentary that earned the highest rating. Llama‑3.3‑70B also maintained perfect formatting; however, its analysis was comparatively superficial and its feedback too terse to guide substantial revision, resulting in a mid‑tier evaluation. DeepSeek‑R1‑70B produced detailed diagnostic notes, yet sporadic formatting lapses and frequent scoring inconsistencies compromised reliability, relegating it to the lowest tier. Consequently, Qwen‑2.5‑72B was selected as the core model for our agent framework.

\section{Conclusions}

By uniting open-source large language models within a purpose-built, multi-agent architecture, our work demonstrates a viable path toward alleviating the twin bottlenecks of proposal preparation and review that presently limit the scientific return of time-dominated astronomical facilities. The Proposal Authoring Agent and Review Agent operate in a closed-loop, “review–refine” paradigm that mirrors real-world TAC workflows, yet executes in minutes rather than weeks. Empirical tests show that proposals iteratively polished by the system score higher on clarity, technical soundness, and scientific merit, while the Review Agent reliably reproduces human acceptance decisions—offering an automated, expert-level second opinion. Taken together, these results indicate that domain-tuned LLM agents can both enhance the quality of investigator submissions and lighten reviewer workloads, paving the way for more efficient, equitable, and impactful transient-focused observing programs in the next generation of astrophysical research.

In future developments, we aim to collaborate with observatories to obtain full-text proposals, both accepted and rejected, along with their corresponding reviewer comments. This will enable the construction of a domain-specific dataset for fine-tuning both the Proposal Authoring Agent and the Review Agent. Additionally, we plan to incorporate observation strategy simulation into the workflow, as the simulated outcomes are essential for evaluating the feasibility and scientific value of a proposal. By integrating this experimental simulation step into the Review Agent’s evaluation process, we allow it to assess proposals not only from a textual and structural standpoint, but also based on the results of parameter-optimized simulations. This multimodal review mechanism would represent a significant step toward achieving fully automated, end-to-end proposal refinement.

\section*{Acknowledgments}
This work was supported by the Strategic Priority Research Program of Chinese Academy of Sciences under Grant XDA0480303.

\printbibliography

@article{grattafiori2024llama,
  title={The {LLaMa} 3 herd of models},
  author={Grattafiori, Aaron and Dubey, Abhimanyu and Jauhri, Abhinav and Pandey, Abhinav and Kadian, Abhishek and Al-Dahle, Ahmad and Letman, Aiesha and Mathur, Akhil and Schelten, Alan and Vaughan, Alex and others},
  journal={arXiv preprint arXiv:2407.21783},
  year={2024}
}

@article{zhang2024grris,
  title={{GRRIS}: A Real-time Intrasite Observation Scheduling Scheme for Distributed Survey Telescope Arrays},
  author={Zhang, Yajie and Yu, Ce and Sun, Chao and Hu, Yi and Shang, Zhaohui and Wei, Jizeng and Yang, Xu},
  journal={The Astronomical Journal},
  volume={168},
  number={5},
  pages={214},
  year={2024},
  publisher={IOP Publishing}
}

@article{williams2025allocating,
  title={Allocating time on scientific platforms in outer space: Evidence from {James Webb} Space Telescope Cycle 1-3 general observer programs},
  author={Williams, Christopher},
  journal={Research Policy},
  volume={54},
  number={5},
  pages={105239},
  year={2025},
  publisher={Elsevier}
}

@article{carpenter2025enhancing,
  title={Enhancing Peer Review in Astronomy: A Machine Learning and Optimization Approach to Reviewer Assignments for {ALMA}},
  author={Carpenter, John M and Corvill{\'o}n, Andrea and Shah, Nihar B},
  journal={Publications of the Astronomical Society of the Pacific},
  volume={137},
  number={3},
  pages={034501},
  year={2025},
  publisher={IOP Publishing}
}

@article{fang2025transientverse,
  title={{TransientVerse}: A Comprehensive Real-Time Alert and Multi-Wavelength Analysis System for Transient Astronomical Events},
  author={Fang, Jian-Hua and Li, Di and Wang, Pei and Chen, Hua-Xi and Wang, Han and Zhou, Deng-Ke and Bao, Qin-Ping and Li, Hai-Yan and Hu, Jing-Jing and Xie, Jin-Tao and others},
  journal={arXiv preprint arXiv:2501.04247},
  year={2025}
}

@article{kerzendorf2020distributed,
  title={Distributed peer review enhanced with natural language processing and machine learning},
  author={Kerzendorf, Wolfgang E and Patat, Ferdinando and Bordelon, Dominic and van de Ven, Glenn and Pritchard, Tyler A},
  journal={Nature Astronomy},
  volume={4},
  number={7},
  pages={711--717},
  year={2020},
  publisher={Nature Publishing Group UK London}
}

@inproceedings{jerabkova2024scientific,
  title={Scientific Text Analysis with Robots applied to observatory proposals},
  author={Jerabkova, Tereza and Boffin, HMJ and Patat, Ferdinando and Dorigo, D and Sogni, F and Primas, F},
  booktitle={Observatory Operations: Strategies, Processes, and Systems X},
  volume={13098},
  pages={509--520},
  year={2024},
  organization={SPIE}
}

@article{idahl2024openreviewer,
  title={{OpenReviewer}: A Specialized Large Language Model for Generating Critical Scientific Paper Reviews},
  author={Idahl, Maximilian and Ahmadi, Zahra},
  journal={arXiv preprint arXiv:2412.11948},
  year={2024}
}

@article{zheng2023judging,
  title={Judging {LLM}-as-a-judge with {MT}-bench and chatbot arena},
  author={Zheng, Lianmin and Chiang, Wei-Lin and Sheng, Ying and Zhuang, Siyuan and Wu, Zhanghao and Zhuang, Yonghao and Lin, Zi and Li, Zhuohan and Li, Dacheng and Xing, Eric and others},
  journal={Advances in Neural Information Processing Systems},
  volume={36},
  pages={46595--46623},
  year={2023}
}

@article{hurst2024gpt,
  title={{GPT}-4o system card},
  author={Hurst, Aaron and Lerer, Adam and Goucher, Adam P and Perelman, Adam and Ramesh, Aditya and Clark, Aidan and Ostrow, AJ and Welihinda, Akila and Hayes, Alan and Radford, Alec and others},
  journal={arXiv preprint arXiv:2410.21276},
  year={2024}
}

@article{li2025astronomical,
  title={An astronomical question answering dataset for evaluating large language models},
  author={Li, Jie and Zhao, Fuyong and Chen, Panfeng and Xie, Jiafu and Zhang, Xiangrui and Li, Hui and Chen, Mei and Wang, Yanhao and Zhu, Ming},
  journal={Scientific Data},
  volume={12},
  number={1},
  pages={447},
  year={2025},
  publisher={Nature Publishing Group UK London}
}

@article{grezes2021building,
  title={Building {astroBERT}, a language model for astronomy \& astrophysics},
  author={Grezes, Felix and Blanco-Cuaresma, Sergi and Accomazzi, Alberto and Kurtz, Michael J and Shapurian, Golnaz and Henneken, Edwin and Grant, Carolyn S and Thompson, Donna M and Chyla, Roman and McDonald, Stephen and others},
  journal={arXiv preprint arXiv:2112.00590},
  year={2021}
}

@inproceedings{devlin2019bert,
  title={{BERT}: Pre-training of deep bidirectional {Transformers} for language understanding},
  author={Devlin, Jacob and Chang, Ming-Wei and Lee, Kenton and Toutanova, Kristina},
  booktitle={Proceedings of the 2019 Conference of the North American chapter of the Association for Computational Linguistics: Human Language Technologies, Volume 1},
  pages={4171--4186},
  year={2019}
}

@article{nguyen2023astrollama,
  title={{AstroLLaMA}: Towards specialized foundation models in astronomy},
  author={Nguyen, Tuan Dung and Ting, Yuan-Sen and Ciuc{\u{a}}, Ioana and O'Neill, Charlie and Sun, Ze-Chang and Jab{\l}o{\'n}ska, Maja and Kruk, Sandor and Perkowski, Ernest and Miller, Jack and Li, Jason and others},
  journal={arXiv preprint arXiv:2309.06126},
  year={2023}
}

@article{ting2025astromlab,
  title={{AstroMLab} 1: Who wins astronomy jeopardy!?},
  author={Ting, Y-S and Nguyen, T Dung and Ghosal, Tirthankar and Pan, Rui and Arora, Hardik and Sun, Zechang and de Haan, Tijmen and Ramachandra, Nesar and Wells, Azton and Madireddy, Sandeep and others},
  journal={Astronomy and Computing},
  volume={51},
  pages={100893},
  year={2025},
  publisher={Elsevier}
}

@inproceedings{pan2024astromlab,
  title={{AstroMLab} 2: {AstroLLaMA-2-70B} Model and Benchmarking Specialised {LLMs} for Astronomy},
  author={Pan, Rui and Nguyen, Tuan Dung and Arora, Hardik and Accomazzi, Alberto and Ghosal, Tirthankar and Ting, Yuan-Sen},
  booktitle={SC24-W: Workshops of the International Conference for High Performance Computing, Networking, Storage and Analysis},
  pages={87--96},
  year={2024},
  organization={IEEE}
}

@article{deastromlab,
  title={{AstroMLab} 3: Achieving GPT-4o Level Performance in Astronomy with a Specialized 8B},
  author={de Haan, Tijmen and Ting, Yuan-Sen and Ghosal, Tirthankar and Nguyen, Tuan Dung},
  journal={Scientific Reports}
}

@article{de2025astromlab,
  title={{AstroMLab} 4: Benchmark-Topping Performance in Astronomy Q\&A with a 70B-Parameter Domain-Specialized Reasoning Model},
  author={de Haan, Tijmen and Ting, Yuan-Sen and Ghosal, Tirthankar and Nguyen, Tuan Dung and Accomazzi, Alberto and Herron, Emily and Lama, Vanessa and Pan, Rui and Wells, Azton and Ramachandra, Nesar},
  journal={arXiv preprint arXiv:2505.17592},
  year={2025}
}

@article{perkowski2024astrollama,
  title={{AstroLLaMA-Chat}: Scaling {AstroLLaMA} with conversational and diverse datasets},
  author={Perkowski, Ernest and Pan, Rui and Nguyen, Tuan Dung and Ting, Yuan-Sen and Kruk, Sandor and Zhang, Tong and O’Neill, Charlie and Jablonska, Maja and Sun, Zechang and Smith, Michael J and others},
  journal={Research Notes of the AAS},
  volume={8},
  number={1},
  pages={7},
  year={2024},
  publisher={The American Astronomical Society}
}

@article{touvron2023llama,
  title={{LLaMa} 2: Open foundation and fine-tuned chat models},
  author={Touvron, Hugo and Martin, Louis and Stone, Kevin and Albert, Peter and Almahairi, Amjad and Babaei, Yasmine and Bashlykov, Nikolay and Batra, Soumya and Bhargava, Prajjwal and Bhosale, Shruti and others},
  journal={arXiv preprint arXiv:2307.09288},
  year={2023}
}

@inproceedings{weng2025cycleresearcher,
title={{CycleResearcher}: Improving Automated Research via Automated Review},
author={Yixuan Weng and Minjun Zhu and Guangsheng Bao and Hongbo Zhang and Jindong Wang and Yue Zhang and Linyi Yang},
booktitle={The Thirteenth International Conference on Learning Representations},
year={2025}}

@article{d2024marg,
  title={{MARG}: Multi-agent review generation for scientific papers},
  author={D'Arcy, Mike and Hope, Tom and Birnbaum, Larry and Downey, Doug},
  journal={arXiv preprint arXiv:2401.04259},
  year={2024}
}

@article{yu2024automated,
  title={Automated peer reviewing in paper sea: Standardization, evaluation, and analysis},
  author={Yu, Jianxiang and Ding, Zichen and Tan, Jiaqi and Luo, Kangyang and Weng, Zhenmin and Gong, Chenghua and Zeng, Long and Cui, Renjing and Han, Chengcheng and Sun, Qiushi and others},
  journal={arXiv preprint arXiv:2407.12857},
  year={2024}
}

@article{zhu2025deepreview,
  title={{DeepReview}: Improving {LLM}-based paper review with human-like deep thinking process},
  author={Zhu, Minjun and Weng, Yixuan and Yang, Linyi and Zhang, Yue},
  journal={arXiv preprint arXiv:2503.08569},
  year={2025}
}

@article{lu2024ai,
  title={The {AI} scientist: Towards fully automated open-ended scientific discovery},
  author={Lu, Chris and Lu, Cong and Lange, Robert Tjarko and Foerster, Jakob and Clune, Jeff and Ha, David},
  journal={arXiv preprint arXiv:2408.06292},
  year={2024}
}

@article{yamada2025ai,
  title={The {AI} scientist-v2: Workshop-level automated scientific discovery via agentic tree search},
  author={Yamada, Yutaro and Lange, Robert Tjarko and Lu, Cong and Hu, Shengran and Lu, Chris and Foerster, Jakob and Clune, Jeff and Ha, David},
  journal={arXiv preprint arXiv:2504.08066},
  year={2025}
}

@article{qwen2024qwen25,
    title={Qwen2.5 Technical Report},
    author={An Yang and Baosong Yang and Beichen Zhang and Binyuan Hui and Bo Zheng and Bowen Yu and Chengyuan Li and Dayiheng Liu and Fei Huang and Haoran Wei and Huan Lin and Jian Yang and Jianhong Tu and Jianwei Zhang and Jianxin Yang and Jiaxi Yang and Jingren Zhou and Junyang Lin and Kai Dang and Keming Lu and Keqin Bao and Kexin Yang and Le Yu and Mei Li and Mingfeng Xue and Pei Zhang and Qin Zhu and Rui Men and Runji Lin and Tianhao Li and Tianyi Tang and Tingyu Xia and Xingzhang Ren and Xuancheng Ren and Yang Fan and Yang Su and Yichang Zhang and Yu Wan and Yuqiong Liu and Zeyu Cui and Zhenru Zhang and Zihan Qiu},
    journal={arXiv preprint arXiv:2412.15115},
    year={2024}
}

@article{deepseekai2025deepseekr1incentivizingreasoningcapability,
      title={{DeepSeek-R1}: Incentivizing Reasoning Capability in LLMs via Reinforcement Learning}, 
      author={DeepSeek-AI},
      year={2025},
      eprint={2501.12948},
      archivePrefix={arXiv},
      primaryClass={cs.CL},
      url={https://arxiv.org/abs/2501.12948}, 
}

\end{document}